\documentclass{article}
\usepackage{amsfonts}
\usepackage{amsmath}

\input{tcilatex}

\begin{document}

\title{Realization of the $N$(odd)-dimensional Quantum Euclidean Space by
Differential Operators\thanks{%
The project supported by National Natural Science Foundation of China under
Grant No. 10075042}}
\author{Yun Li\thanks{%
Email: yunlee@mail.ustc.edu.cn} and Sicong Jing \\
\textit{Department of Modern Physics, }\\
\textit{\ University of Science and Technology of China,}\\
\textit{\ Hefei, Anhui 230026, P.R.China }}
\maketitle

\begin{abstract}
The quantum Euclidean space $%
\mathbb{R}
_{q}^{N}$ is a kind of noncommutative space which is obtained from ordinary
Euclidean space $%
\mathbb{R}
^{N}$ by deformation with parameter $q$. When $N$ is odd, the structure of
this space is similar to $%
\mathbb{R}
_{q}^{3}$. Motivated by realization of $%
\mathbb{R}
_{q}^{3}$ by differential operators in $%
\mathbb{R}
^{3}$, we give such realization for $%
\mathbb{R}
_{q}^{5}$ and $%
\mathbb{R}
_{q}^{7}$ cases and generalize our results to $%
\mathbb{R}
_{q}^{N}$ ($N$ odd) in this paper, that is, we show that the algebra of $%
\mathbb{R}
_{q}^{N}$ can be realized by differential operators acting on $C^{\infty }$
functions on undeformed space $%
\mathbb{R}
^{N}$.

\end{abstract}

\section{Introduction}

Deformation plays an important role in the development of theoretical
models. Indeed the passage from Galilei relativity to that of Lorentz as
well as transition from classical to quantum mechanics can both be regarded
as deformations associated with the dimensional deformation parameter $c$
(the velocity of light) and
h{\hskip-.2em}\llap{\protect\rule[1.1ex]{.325em}{.1ex}}{\hskip.2em}
(the Plank constant) respectively. With the intensive study of quantum
groups and their associated noncommutative geometry, one builds the
Euclidean quantum space $%
\mathbb{R}
_{q}^{N}$ related to orthogonal quantum group $SO_{q}(N)$. In Ref.\cite{1}
and \cite{2}, some physical systems were constructed based on those
Euclidean quantum spaces. So with the view of the historical examples in
mind one can look on those physical systems as a deformation of ordinary
physics with deformation parameter $q$.

In order to learn more deformed physical system based on\ the
three-dimensional Euclidean quantum space $%
\mathbb{R}
_{q}^{3}$, recently the structure of $%
\mathbb{R}
_{q}^{3}$ was discussed in detail\cite{3} and in Ref.\cite{4} the coordinate
elements of $%
\mathbb{R}
_{q}^{3}$ were represented by differential operators in undeformed Euclidean
space $%
\mathbb{R}
^{3}$. This differential realization is not a novel approach. In fact the
method of realizing the generators of universal deformed quantum groups in
terms of standard differential operators was discussed thoroughly in Ref.%
\cite{5}.

For the general algebra of $%
\mathbb{R}
_{q}^{N}$ it was found that the cases $N$ even and $N$ odd are somewhat
different. When $N$ is odd the geometrical structure is similar to $%
\mathbb{R}
_{q}^{3}$ case. When, on the other hand, $N$ is even, one should add an
additional component to the algebra of $%
\mathbb{R}
_{q}^{N}$ \cite{6}. Also in Ref.\cite{7} for the sake of the different
parity of $N$ the Euclidean quantum space $%
\mathbb{R}
_{q}^{N}$ was sorted as $B_{N}$ ($N$ odd) and $D_{N}$ ($N$ even) separately
and, for $B_{N}$ case, the commutation relations for coordinates, as the
so-called algebra of the Euclidean quantum space $%
\mathbb{R}
_{q}^{N}$, are given by

\begin{eqnarray}
X_{i}X_{j} &=&qX_{j}X_{i},\text{ \ \ }-n\leq i<j\leq n,i\neq -j,  \label{1}
\\
X_{j}X_{-j} &=&X_{-j}X_{j}+q\lambda \sum_{k=-n}^{j-1}q^{\rho _{j}-\rho
_{k}}X_{k}X_{-k}-\frac{\lambda q^{\rho _{j}+1}}{1+q^{2n-1}}%
\sum_{k=-n}^{n}q^{-\rho _{k}}X_{k}X_{-k},\text{ }  \notag
\end{eqnarray}%
where $n=\frac{N-1}{2},$ $\lambda =q-q^{-1}$ and $\rho _{-i}=i-\frac{1}{2}%
,\rho _{0}=0,\rho _{i}=-i+\frac{1}{2}$ $(i>0)$. The coordinates $X_{i}$
satisfy the following conjugation relation

\begin{equation}
\overline{X}_{i}=q^{\rho _{-i}}X_{-i}\text{ .}  \label{2}
\end{equation}

In this paper we restrict our discussion on such $B_{N}$ case and propose to
give a kind of specific realization of the $N$(odd)-dimensional Euclidean
quantum space by differential operators in $%
\mathbb{R}
^{N}$. We extend the approach developed in Ref.\cite{4} and the result we
shall see is somewhat interesting. In section 2 we give a brief review of $%
N=3$ case, which is discussed in Ref.\cite{4}. In section 3 we discuss the
case of $N=5$ and $7$. In section 4 we will generalize our results to the
general $N$(odd) case.

\section{Review of $%
\mathbb{R}
_{q}^{3}$ case}

If one set $n=1$ in Eq.(1), one obtains a $SO_{q}(3)$-module algebra. This
is the algebra of the Euclidean quantum space $%
\mathbb{R}
_{q}^{3}$ with definite relations:

\begin{eqnarray}
X_{-1}X_{0} &=&qX_{0}X_{-1},  \label{3} \\
X_{0}X_{1} &=&qX_{1}X_{0},  \notag \\
X_{1}X_{-1} &=&X_{-1}X_{1}+(q^{1/2}-q^{-1/2})X_{0}X_{0}\text{ .}  \notag
\end{eqnarray}%
From (2), it is natural to have the conjugation relation of the generators
of the above algebra based on $%
\mathbb{R}
_{q}^{3}$:

\begin{equation}
\overline{X}_{1}=q^{1/2}X_{-1}\text{ .}  \label{4}
\end{equation}%
The next purpose is to represent the algebra in terms of differential
operators acting on $%
\mathbb{R}
^{3}$. For convenience we use polar coordinates $(r,\theta ,\varphi )$ and
define $\xi =\cos \theta $. An operator that will play a major role in our
discussion is:

\begin{equation}
\Lambda _{\xi }=\tfrac{1}{2}(\xi \tfrac{\partial }{\partial \xi }+\tfrac{%
\partial }{\partial \xi }\xi )=\xi \tfrac{\partial }{\partial \xi }+\tfrac{1%
}{2}\text{ .}  \label{5}
\end{equation}%
When acting on $L^{2}$-functions in the common domain of $\Lambda _{\xi }$
and $\Lambda _{\xi }^{\ast }$,

\begin{equation}
\Lambda _{\xi }^{\ast }=-\Lambda _{\xi }  \label{6}
\end{equation}%
will hold. The property of $\Lambda _{\xi }$ is followed by

\begin{eqnarray}
\left[ \Lambda _{\xi },\xi \right] &=&\xi ,  \label{7} \\
e^{\alpha \Lambda _{\xi }}\xi e^{-\alpha \Lambda _{\xi }} &=&e^{\alpha }\xi
\text{ .}  \notag
\end{eqnarray}%
Now we make an ansatz:

\begin{eqnarray}
X_{0} &=&r\xi ,  \label{8} \\
X_{1} &=&rf(\xi )e^{-\alpha \Lambda _{\xi }},  \notag \\
X_{-1} &=&rg(\xi )e^{\beta \Lambda _{\xi }}\text{ .}  \notag
\end{eqnarray}%
Here, $f(\xi )$ and $g(\xi )$ are complex functions acting on $%
\mathbb{C}
^{1}$. From the first two equations of (3), it is easy to verify

\begin{equation}
e^{\alpha }=e^{\beta }=q  \label{9}
\end{equation}%
and from the third equation:

\begin{equation}
f(\xi )g(q^{-1}\xi )-f(q\xi )g(\xi )=(q^{1/2}-q^{-1/2})\xi ^{2}\text{ .}
\label{10}
\end{equation}%
If we define

\begin{equation}
\Phi (\xi )=f(\xi )g(q^{-1}\xi ),  \label{11}
\end{equation}%
Eq.(10) becomes:

\begin{equation}
\Phi (\xi )-\Phi (q\xi )=(q^{1/2}-q^{-1/2})\xi ^{2}  \label{12}
\end{equation}%
with the solution:

\begin{equation}
\Phi (\xi )=\Phi (0)-\frac{q^{-1/2}}{1+q}\xi ^{2}\text{ .}  \label{13}
\end{equation}%
Next by identifying the radius $r$ with the invariant length in $%
\mathbb{R}
_{q}^{3}$:

\begin{equation}
r^{2}=R^{2}\equiv q^{-1/2}X_{-1}X_{1}+X_{0}X_{0}+q^{1/2}X_{1}X_{-1},
\label{14}
\end{equation}%
we can determine $\Phi (0)$:

\begin{equation}
\Phi (0)=\frac{q^{1/2}}{1+q}\text{ .}  \label{15}
\end{equation}%
With the conjugation relation(4), the following equation is obtained

\begin{equation}
g(\xi )=q^{-1/2}\overline{f(q\xi )}  \label{16}
\end{equation}%
and, as a consequence of the definition(11) of $\Phi (\xi )$, we have

\begin{equation}
\Phi (\xi )=q^{-1/2}\left\vert f(\xi )\right\vert ^{2}\text{ .}  \label{17}
\end{equation}%
Combining (13), (15)\ and (17)\ leads to the following result

\begin{eqnarray}
X_{0} &=&r\xi ,  \label{18} \\
X_{1} &=&re^{i\varphi }\sqrt{\frac{1}{1+q^{-1}}-\frac{1}{1+q}\xi ^{2}}%
q^{-\Lambda _{\xi }},  \notag \\
X_{-1} &=&re^{-i\varphi }\sqrt{\frac{q^{-1}}{1+q^{-1}}-\frac{q}{1+q}\xi ^{2}}%
q^{\Lambda _{\xi }}\text{ .}  \notag
\end{eqnarray}

\section{$%
\mathbb{R}
_{q}^{5}$ and $%
\mathbb{R}
_{q}^{7}$ case}

When we set $n=2$ in Eq.(1), we will have the $SO_{q}(5)$-module algebra:

\begin{eqnarray}
X_{i}X_{j} &=&qX_{j}X_{i},\text{ \ \ }-2\leq i<j\leq 2,i\neq -j,  \label{19}
\\
X_{1}X_{-1} &=&X_{-1}X_{1}+(q^{1/2}-q^{-1/2})X_{0}X_{0},  \notag \\
X_{2}X_{-2} &=&X_{-2}X_{2}+qX_{1}X_{-1}-q^{-1}X_{-1}X_{1},  \notag
\end{eqnarray}%
and the corresponding conjugation relations of the generators are given by

\begin{eqnarray}
\overline{X}_{1} &=&q^{1/2}X_{-1},  \label{20} \\
\overline{X}_{2} &=&q^{3/2}X_{-2}\text{ .}  \notag
\end{eqnarray}%
To represent the algebra(19) in terms of differential operators acting on
the Euclidean space $%
\mathbb{R}
^{5}$, we also use polar coordinates $(r,\theta _{1},\theta _{2},\varphi
_{1},\varphi _{2})$ of $%
\mathbb{R}
^{5}$ and define $\xi _{i}=\cos \theta _{i}$ $(i=1,2)$. Like the ansatz made
in Eq.(8) when dealing with $%
\mathbb{R}
_{q}^{3}$ case and after considering the explicit expression(18),
we assume that the generators $X_{i}$ have the following form:

\begin{eqnarray}
X_{0} &=&r\xi _{1},  \label{21} \\
X_{1} &=&re^{i\varphi _{1}}f_{1}(\xi _{1},\xi _{2})e^{-\alpha _{1}\Lambda
_{\xi _{1}}},  \notag \\
X_{-1} &=&re^{-i\varphi _{1}}g_{1}(\xi _{1},\xi _{2})e^{\beta _{1}\Lambda
_{\xi _{1}}},  \notag \\
X_{2} &=&re^{i\varphi _{2}}f_{2}(\xi _{1},\xi _{2})e^{-\alpha _{2}(\Lambda
_{\xi _{1}}+\Lambda _{\xi _{2}})},  \notag \\
X_{-2} &=&re^{-i\varphi _{2}}g_{2}(\xi _{1},\xi _{2})e^{\beta _{2}(\Lambda
_{\xi _{1}}+\Lambda _{\xi _{2}})}\text{ .}  \notag
\end{eqnarray}%
Here, $f_{1,2}(\xi _{1},\xi _{2})$ and $g_{1,2}(\xi _{1},\xi _{2})$ are
absolutely real functions acting on $%
\mathbb{R}
^{2}$ because we write the complex parts into exponential forms in the above
expressions. From the first equation of (19), the following

\begin{equation}
e^{\alpha _{i}}=e^{\beta _{i}}=q,\text{ \ }(i=1,2)  \label{22}
\end{equation}%
also holds. Simultaneously some confining conditions related to the
functions $f_{1,2}(\xi _{1},\xi _{2})$ and $g_{1,2}(\xi _{1},\xi _{2})$ are
obtained. We list them below%
\begin{eqnarray}
g_{1}(q\xi _{1},q\xi _{2})g_{2}(\xi _{1},\xi _{2}) &=&qg_{1}(\xi _{1},\xi
_{2})g_{2}(q\xi _{1},\xi _{2}),  \notag \\
f_{1}(q\xi _{1},q\xi _{2})g_{2}(\xi _{1},\xi _{2}) &=&qf_{1}(\xi _{1},\xi
_{2})g_{2}(q^{-1}\xi _{1},\xi _{2}),  \label{23} \\
g_{1}(\xi _{1},\xi _{2})f_{2}(q\xi _{1},\xi _{2}) &=&qg_{1}(q^{-1}\xi
_{1},q^{-1}\xi _{2})f_{2}(\xi _{1},\xi _{2}),  \notag \\
f_{1}(\xi _{1},\xi _{2})f_{2}(q^{-1}\xi _{1},\xi _{2}) &=&qf_{1}(q^{-1}\xi
_{1},q^{-1}\xi _{2})f_{2}(\xi _{1},\xi _{2}).  \notag
\end{eqnarray}%
From the second equation of (19), we have

\begin{equation}
f_{1}(\xi _{1},\xi _{2})g_{1}(q^{-1}\xi _{1},\xi _{2})-f_{1}(q\xi _{1},\xi
_{2})g_{1}(\xi _{1},\xi _{2})=(q^{1/2}-q^{-1/2})\xi _{1}^{2}\text{ .}
\label{24}
\end{equation}%
With the definition:

\begin{equation}
\Phi _{1}(\xi _{1},\xi _{2})=f_{1}(\xi _{1},\xi _{2})g_{1}(q^{-1}\xi
_{1},\xi _{2}),  \label{25}
\end{equation}%
this equation becomes:

\begin{equation}
\Phi _{1}(\xi _{1},\xi _{2})-\Phi _{1}(q\xi _{1},\xi
_{2})=(q^{1/2}-q^{-1/2})\xi _{1}^{2}  \label{26}
\end{equation}%
and has the solution:

\begin{equation}
\Phi _{1}(\xi _{1},\xi _{2})=\Phi _{1}(0,0)-\frac{q^{-1/2}}{1+q}\xi
_{1}^{2}+B_{1}\xi _{2}^{2},  \label{27}
\end{equation}%
in which $\Phi _{1}(0,0)$ and $B_{1}$ need to be determined further. The
third equation of (19) leads to the results:

\begin{eqnarray}
&&f_{2}(\xi _{1},\xi _{2})g_{2}(q^{-1}\xi _{1},q^{-1}\xi _{2})  \label{28} \\
&=&f_{2}(q\xi _{1},q\xi _{2})g_{2}(\xi _{1},\xi _{2})+qf_{1}(\xi _{1},\xi
_{2})g_{1}(q^{-1}\xi _{1},\xi _{2})-q^{-1}f_{1}(q\xi _{1},\xi _{2})g_{1}(\xi
_{1},\xi _{2})\text{ .}  \notag
\end{eqnarray}%
By defining

\begin{equation}
\Phi _{2}(\xi _{1},\xi _{2})=f_{2}(\xi _{1},\xi _{2})g_{2}(q^{-1}\xi
_{1},q^{-1}\xi _{2}),  \label{29}
\end{equation}%
we have the following

\begin{equation}
\Phi _{2}(\xi _{1},\xi _{2})=\Phi _{2}(q\xi _{1},q\xi _{2})+q\Phi _{1}(\xi
_{1},\xi _{2})-q^{-1}\Phi _{1}(q\xi _{1},\xi _{2}).  \label{30}
\end{equation}%
Substituting the solution(27) of $\Phi _{1}(\xi _{1},\xi _{2})$ into the
above equation, consequently we have the solution of $\Phi _{2}(\xi _{1},\xi
_{2}):$

\begin{equation}
\Phi _{2}(\xi _{1},\xi _{2})=\Phi _{2}(0,0)-q^{-1}B_{1}\xi _{2}^{2}
\label{31}
\end{equation}%
and also we can determine the value of $\Phi _{1}(0,0)$ satisfying

\begin{equation}
\Phi _{1}(0,0)=0\text{.}  \label{32}
\end{equation}%
Next also by identifying radius $r$ with the invariant length in $%
\mathbb{R}
_{q}^{5}$:

\begin{equation}
r^{2}=R^{2}\equiv
q^{-3/2}X_{-2}X_{2}+q^{-1/2}X_{-1}X_{1}+X_{0}X_{0}+q^{1/2}X_{1}X_{-1}+q^{3/2}X_{2}X_{-2}
\label{33}
\end{equation}%
and after a careful calculation, $\Phi _{2}(0,0)$ can be determined:

\begin{equation}
\Phi _{2}(0,0)=\frac{1}{q^{3/2}+q^{-3/2}}\text{ .}  \label{34}
\end{equation}%
So we have

\begin{eqnarray}
\Phi _{1}(\xi _{1},\xi _{2}) &=&B_{1}\xi _{2}^{2}-\frac{q^{-1/2}}{1+q}\xi
_{1}^{2},  \label{35} \\
\Phi _{2}(\xi _{1},\xi _{2}) &=&\frac{1}{q^{3/2}+q^{-3/2}}-q^{-1}B_{1}\xi
_{2}^{2}  \notag
\end{eqnarray}%
only with parameter $B_{1}$ needing to be determined further. By applying
the conjugation relations(20), we derive the following equations

\begin{eqnarray}
\Phi _{1}(\xi _{1},\xi _{2}) &=&q^{-1/2}f_{1}^{2}(\xi _{1},\xi _{2}),
\label{36} \\
\Phi _{2}(\xi _{1},\xi _{2}) &=&q^{-3/2}f_{2}^{2}(\xi _{1},\xi _{2})\text{ .}
\notag
\end{eqnarray}%
With (35) and (36) it is easy to obtain

\begin{eqnarray}
X_{0} &=&r\xi _{1},  \label{37} \\
X_{1} &=&re^{i\varphi _{1}}\sqrt{q^{1/2}B_{1}\xi _{2}^{2}-\frac{1}{1+q}\xi
_{1}^{2}}q^{-\Lambda _{\xi _{1}}},\text{\ }  \notag \\
\text{\ }X_{-1} &=&re^{-i\varphi _{1}}\sqrt{q^{-1/2}B_{1}\xi _{2}^{2}-\frac{q%
}{1+q}\xi _{1}^{2}}q^{\Lambda _{\xi _{1}}},  \notag \\
X_{2} &=&re^{i\varphi _{2}}\sqrt{\frac{1}{1+q^{-3}}-q^{1/2}B_{1}\xi _{2}^{2}}%
q^{-(\Lambda _{\xi _{1}}+\Lambda _{\xi _{2}})},  \notag \\
\text{ \ }X_{-2} &=&re^{-i\varphi _{2}}\sqrt{\frac{q^{-3}}{1+q^{-3}}%
-q^{-1/2}B_{1}\xi _{2}^{2}}q^{\Lambda _{\xi _{1}}+\Lambda _{\xi _{2}}}\text{.%
}  \notag
\end{eqnarray}%
By setting

\begin{equation}
B_{1}=\frac{q^{-1/2}}{1+q^{3}},  \label{38}
\end{equation}%
the above expression can have a uniform expression with (18). So the
generators $X_{i}$ of the $SO_{q}(5)$-module algebra can be written as

\begin{eqnarray}
X_{0} &=&r\xi _{1},  \notag \\
X_{1} &=&re^{i\varphi _{1}}\sqrt{\frac{1}{1+q^{3}}\xi _{2}^{2}-\frac{1}{1+q}%
\xi _{1}^{2}}q^{-\Lambda _{\xi _{1}}},  \notag \\
X_{-1} &=&re^{-i\varphi _{1}}\sqrt{\frac{q^{-1}}{1+q^{3}}\xi _{2}^{2}-\frac{q%
}{1+q}\xi _{1}^{2}}q^{\Lambda _{\xi _{1}}},  \label{39} \\
X_{2} &=&re^{i\varphi _{2}}\sqrt{\frac{1}{1+q^{-3}}-\frac{1}{1+q^{3}}\xi
_{2}^{2}}q^{-(\Lambda _{\xi _{1}}+\Lambda _{\xi _{2}})},  \notag \\
X_{-2} &=&re^{-i\varphi _{2}}\sqrt{\frac{q^{-3}}{1+q^{-3}}-\frac{q^{-1}}{%
1+q^{3}}\xi _{2}^{2}}q^{\Lambda _{\xi _{1}}+\Lambda _{\xi _{2}}}\text{.}
\notag
\end{eqnarray}%
The above expressions are obviously consistent with the confining conditions
metioned in Eq.(23). Our purpose is to represent the algebra of the $N$%
(odd)-dimensional Euclidean quantum space in terms of differential operators
acting on $%
\mathbb{R}
^{N}$, so if we obtain the representation in $%
\mathbb{R}
_{q}^{7}$ case again, we can find the intrinsic rule of expression forms and
generalize our results to the $N$(odd) case. For $N=7$ case, the algebra
relations are given by

\begin{eqnarray}
X_{i}X_{j} &=&qX_{i}X_{j},\text{ \ \ }-3\leq i<j\leq 3,i\neq -j,  \notag \\
X_{1}X_{-1} &=&X_{-1}X_{1}+(q^{1/2}-q^{-1/2})X_{0}X_{0},  \label{40} \\
X_{2}X_{-2} &=&X_{-2}X_{2}+qX_{1}X_{-1}-q^{-1}X_{-1}X_{1},  \notag \\
X_{3}X_{-3} &=&X_{-3}X_{3}+qX_{2}X_{-2}-q^{-1}X_{-2}X_{2}  \notag
\end{eqnarray}%
with the conjugation relations of the generators:

\begin{eqnarray}
\overline{X}_{1} &=&q^{1/2}X_{-1},  \notag \\
\overline{X}_{2} &=&q^{3/2}X_{-2},  \label{41} \\
\overline{X}_{3} &=&q^{5/2}X_{-3}\text{.}  \notag
\end{eqnarray}%
Following the same procedure as we obtain expression(39), we have the
following

\begin{eqnarray}
X_{0} &=&r\xi _{1},  \notag \\
X_{1} &=&re^{i\varphi _{1}}\sqrt{\frac{1}{1+q^{3}}\xi _{2}^{2}-\frac{1}{1+q}%
\xi _{1}^{2}}q^{-\Lambda _{\xi _{1}}},  \notag \\
X_{-1} &=&re^{-i\varphi _{1}}\sqrt{\frac{q^{-1}}{1+q^{3}}\xi _{2}^{2}-\frac{q%
}{1+q}\xi _{1}^{2}}q^{\Lambda _{\xi _{1}}},  \notag \\
X_{2} &=&re^{i\varphi _{2}}\sqrt{\frac{1}{1+q^{5}}\xi _{3}^{2}-\frac{1}{%
1+q^{3}}\xi _{2}^{2}}q^{-(\Lambda _{\xi _{1}}+\Lambda _{\xi _{2}})},
\label{42} \\
X_{-2} &=&re^{-i\varphi _{2}}\sqrt{\frac{q^{-3}}{1+q^{5}}\xi _{3}^{2}-\frac{%
q^{-1}}{1+q^{3}}\xi _{2}^{2}}q^{\Lambda _{\xi _{1}}+\Lambda _{\xi _{2}}},
\notag \\
X_{3} &=&re^{i\varphi _{3}}\sqrt{\frac{1}{1+q^{-5}}-\frac{1}{1+q^{5}}\xi
_{3}^{2}}q^{-(\Lambda _{\xi _{1}}+\Lambda _{\xi _{2}}+\Lambda _{\xi _{3}})},
\notag \\
X_{-3} &=&re^{-i\varphi _{3}}\sqrt{\frac{q^{-5}}{1+q^{-5}}-\frac{q^{-3}}{%
1+q^{5}}\xi _{3}^{2}}q^{\Lambda _{\xi _{1}}+\Lambda _{\xi _{2}}+\Lambda
_{\xi _{3}}}\text{.}  \notag
\end{eqnarray}%
Also here $(r,\theta _{1},\theta _{2},\theta _{3},\varphi _{1},\varphi
_{2},\varphi _{3})$ represent polar coordinates of the Euclidean space $%
\mathbb{R}
^{7}$ and the definitions of $\xi _{i}$ and $\Lambda _{\xi _{i}}$ are same
as (18) and (39).

\section{$%
\mathbb{R}
_{q}^{N}$ ($N$ odd) case}

Now we deal with the algebra of $%
\mathbb{R}
_{q}^{N}$ ($N$ odd) and give the differential realization of it. After
speculating the realization \ in $N=3,5$ and $7$ cases, it is not difficult
to find that for the general $N$ (odd) case the realization can be expressed
as

\begin{eqnarray}
X_{0} &=&r\xi _{1},  \label{43} \\
X_{i} &=&re^{i\varphi _{i}}\sqrt{\frac{1}{1+q^{2i+1}}\xi _{i+1}^{2}-\frac{1}{%
1+q^{2i-1}}\xi _{i}^{2}}q^{-\sum_{m=1}^{i}\Lambda _{\xi _{m}}}\text{\ ,}
\notag \\
X_{-i} &=&re^{-i\varphi _{i}}\sqrt{\frac{q^{-2i+1}}{1+q^{2i+1}}\xi
_{i+1}^{2}-\frac{q^{-2i+3}}{1+q^{2i-1}}\xi _{i}^{2}}q^{\sum_{m=1}^{i}\Lambda
_{\xi _{m}}},\text{ \ \ }1\leq i\leq n-1,n=\frac{N-1}{2},  \notag \\
X_{n} &=&re^{i\varphi _{n}}\sqrt{\frac{1}{1+q^{-2n+1}}-\frac{1}{1+q^{2n-1}}%
\xi _{n}^{2}}q^{-\sum_{m=1}^{n}\Lambda _{\xi _{m}}},  \notag \\
X_{-n} &=&re^{-i\varphi _{n}}\sqrt{\frac{q^{-2n+1}}{1+q^{-2n+1}}-\frac{%
q^{-2n+3}}{1+q^{2n-1}}\xi _{n}^{2}}q^{\sum_{m=1}^{n}\Lambda _{\xi _{m}}}%
\text{.}  \notag
\end{eqnarray}%
We also assume that $(r,\theta _{1},\cdots ,\theta _{i},\cdots ,\theta
_{n},\varphi _{1},\cdots ,\varphi _{i},\cdots \varphi _{n})$ are polar
coordinates of the Euclidean space $%
\mathbb{R}
^{N}$ and have $\xi _{i}=\cos \theta _{i},\Lambda _{\xi }=\xi \tfrac{%
\partial }{\partial \xi }+\tfrac{1}{2}$ in the above expression.

Next we will verify that the above expressions are definitely a kind of
differential realization for the general $N$ (odd) case. One can rewrite the
last line of the commutative relations(1) in a more simple form\cite{8}:

\begin{equation}
X_{s}X_{-s}=X_{-s}X_{s}+\lambda q^{s-3/2}(\sum\limits_{i=1}^{s-1}q^{\rho
_{i}}X_{-i}X_{i}+\frac{q}{1+q}X_{0}X_{0}),\text{ }s>0  \label{44}
\end{equation}%
For $s=1$, it becomes

\begin{equation}
X_{1}X_{-1}=X_{-1}X_{1}+(q^{1/2}-q^{-1/2})X_{0}X_{0},  \label{45}
\end{equation}%
For $s>1$, this relation is equivalent to the following one

\begin{equation}
X_{s}X_{-s}=X_{-s}X_{s}+qX_{s-1}X_{-s+1}-q^{-1}X_{-s+1}X_{s-1}  \label{46}
\end{equation}%
So we can rewrite the algebra of $%
\mathbb{R}
_{q}^{N}$ ($N$ odd)(1) as following

\begin{eqnarray}
X_{i}X_{j} &=&qX_{j}X_{i},\text{ \ \ }-n\leq i<j\leq n,i\neq -j,  \label{47}
\\
X_{1}X_{-1} &=&X_{-1}X_{1}+(q^{1/2}-q^{-1/2})X_{0}X_{0},  \notag \\
X_{s}X_{-s} &=&X_{-s}X_{s}+qX_{s-1}X_{-s+1}-q^{-1}X_{-s+1}X_{s-1},\text{ }%
1<s\leq n  \notag
\end{eqnarray}%
Now substituting the expressions(43) into the above relations we will find
that the right-hand side of each line in the relations is equal to the
left-hand one. So Eq.(43) is a possible realization by standard differential
operators in $%
\mathbb{R}
^{N}$. To make sense, we take the limit $q\rightarrow 1$ in (43) and find
that $X_{i}$ $($\ $-n\leq i\leq n)$ become linearly independent. In this
limit the deformed quantum space $%
\mathbb{R}
_{q}^{N}$ reduces into the undeformed $%
\mathbb{R}
^{N}$ and the generators $X_{i}$ could be a set of basis of $%
\mathbb{R}
^{N}$.

\end{document}